\documentclass[prl,floatfix,reprint,superscriptaddress]{revtex4-2}

\usepackage{amsmath}
\usepackage{amssymb}
\usepackage{bm}
\usepackage{color}
\usepackage{graphicx}
\usepackage{placeins}
\usepackage{hyperref}

\newcommand\pdag{{\vphantom\dag}}
\renewcommand\Im{\operatorname{Im}}
\newcommand\sgn{\operatorname{sgn}}

\begin{document}

\title{Minimal chiral \emph{p}-wave superconductor}

\author{Fabian Hassler}
\affiliation{Institute for Quantum Information, RWTH Aachen University, 52056 Aachen, Germany}
\author{Michael Wimmer}
\affiliation{QuTech and Kavli Institute of Nanoscience, Delft University of Technology, 2600 GA Delft, The Netherlands}
\date{November 2026}
\begin{abstract}
  Chiral $p$-wave superconductors are among the most intriguing topological phases: their vortices bind Majorana zero modes and their edges carry chiral Majorana modes, yet no material has been unambiguously identified as one.
  Here, we carry the bottom-up approach of Kitaev chains to two dimensions, proposing a minimal chiral $p$-wave superconductor from a square array of quantum dots coupled via Andreev dots.
  Rashba spin-orbit coupling in the tunnel barriers provides the $p$-wave winding, and a Zeeman field renders the dots spinless.
  At a sweet spot, the edge states are confined to the outermost row, so that a chiral thermal conductance and a vortex-bound Majorana zero mode with sub-percent splitting emerge already in $3\times3$ and $5\times5$ arrays.
  Our setup is realizable with existing InAs and Ge quantum-dot arrays.
\end{abstract}
\maketitle

\emph{Introduction}---
A \emph{chiral} $p$-wave superconductor pairs spinless fermions in two dimensions (2D) in the $p_x\pm i p_y$ channel, which results in a fully gapped topological phase~\cite{read2000,alicea2012}.
Each edge carries one chiral Majorana mode, and each vortex with flux $\Phi_0=h/(2e)$ binds a \emph{Majorana zero mode}~\cite{ivanov2001}, so that the vortices are Ising anyons whose braiding is topologically protected~\cite{nayak2008}.
Pairing of composite fermions in the same channel underlies the Pfaffian state proposed for the even-denominator fractional quantum Hall effect~\cite{moore1991,read2000}, where a half-integer thermal Hall conductance has established non-Abelian order~\cite{banerjee2018,dutta2022}.

Yet no material has been conclusively established as a chiral $p$-wave superconductor~\cite{kallin2016}.
For Sr$_2$RuO$_4$, long the leading candidate~\cite{rice1995}, the order parameter remains unsettled~\cite{maeno2024}, while UTe$_2$ and rhombohedral graphene show first signatures of chiral superconductivity~\cite{jiao2020,aoki2022,han2025}.
Synthetic routes via proximitized topological-insulator surfaces~\cite{fu2008}, quantum anomalous Hall layers~\cite{qi2010}, Rashba wires~\cite{lutchyn2010,oreg2010} and Ge hole gases~\cite{coppini2026} are promising, yet none has delivered a hard gap with well-separated Majorana modes~\cite{flensberg2021}.

For topological superconductivity in one dimension, a bottom-up approach has recently emerged: assembling Kitaev chains from quantum dots coupled through superconductors~\cite{sau2012,leijnse2012,fulga2013}.
Andreev bound states in hybrid superconducting regions can mediate the required cotunneling and crossed Andreev reflection processes for the Kitaev chains~\cite{liu2022}.
At the Kitaev sweet spot, spatially separated Majorana modes emerge even in a two-dot chain~\cite{leijnse2012}.
Such minimal chains have been realized in hybrid nanowires and 2D electron gases~\cite{dvir2023,tenhaaf2024}, and extensions to three sites have demonstrated increased stability and an interior excitation gap~\cite{bordin2025,tenhaaf2025}.
The key advance of this approach is the control over every dot level and every coupling, which compensates for disorder and reveals topological features already in systems of a few dots.
\begin{figure}[bt]
  \centering
  \includegraphics{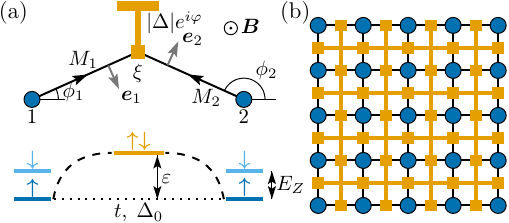}
  \caption{%
  (a) Andreev link: the quantum dots 1, 2 (blue circles) are coupled via an Andreev dot (orange square) with energy level $\xi$, which is proximitized by a superconductor with pairing $|\Delta|e^{i\varphi}$.
  The tunneling from dot $j$ to the Andreev dot in the direction $\phi_j$ is described by the matrix $M_j$, which includes the spin-orbit interaction with direction $\bm e_j$ (gray arrows); an out-of-plane magnetic field $\bm B$ renders the dots effectively spinless.
  Bottom: level scheme.
  The Zeeman-split levels of the quantum dots (blue) are tunnel coupled to the spin-degenerate Andreev level at excitation energy $\varepsilon$ (orange).
  Integrating out the Andreev level yields an effective link (dotted) between the spin-${\uparrow}$ levels with hopping $t$ and pairing $\Delta_0$.
  (b) Square lattice of dots that implements the minimal chiral $p$-wave superconductor in InAs ($\eta=-1$) or in a Ge hole gas ($\eta=+1$).
  The superconducting islands of the Andreev dots are connected by superconducting wires (orange lines) to form a single condensate with a common phase.
  A vortex  centered on one of the quantum dots traps a Majorana zero mode (not shown).
  }\label{fig:setup}
\end{figure}

In this Letter, we propose a minimal chiral $p$-wave superconductor (CPWSC) that carries this approach to 2D, realized as a square array of quantum dots coupled through proximitized Andreev dots [Fig.~\ref{fig:setup}(a)].
No exotic material is needed, as the chirality stems from the Rashba spin-orbit coupling in the tunnel barriers, which makes the phase of crossed Andreev reflection depend on the bond direction, while a Zeeman field selects a single spin species.
We show that, at the 2D analog of the Kitaev sweet spot, chiral edge transport and a resolvable vortex Majorana zero mode appear in arrays of $3\times 3$ or $5\times 5$ dots.
In the following, we first develop the tight-binding model and locate the sweet spot, and then show its implementation in InAs and Ge; finally, we discuss chiral edge transport, vortex Majorana zero modes, and then comment on the robustness to disorder.

\emph{Minimal chiral $p$-wave model}--- The model for the minimal CPWSC is realized as a tight-binding model on a square lattice \footnote{%
`Minimal' refers to tight-binding models with nearest-neighbor couplings; only such local couplings define a 2D geometry and thus an edge that can carry a chiral Majorana mode.
Like the minimal Kitaev chain, the array can be scaled up to larger sizes.}.
Each lattice site represents a quantum dot coupled to its neighbors via an Andreev dot, i.e., a quantum dot that is proximity-coupled to a superconductor.
The targeted tight-binding Hamiltonian
\begin{align}\label{eq:cp}
  H_\text{cp} &= -t \sum_{\langle \bm j,\bm k\rangle} ( c^\dag_{\bm k} c^\pdag_{\bm j} + \text{H.c.}) - \mu \sum_{\bm j} c^\dag_{\bm j} c^\pdag_{\bm j} \nonumber\\
	      &\quad + \Delta_0 \sum_{\bm j} \Bigl[ ( c^\dag_{\bm j+ \hat{\bm x} }  + i \eta c^\dag_{\bm j+ \hat{\bm y}} ) c^\dag_{\bm j} + \text{H.c.} \Bigr]
\end{align}
implements a CPWSC with $p_x + i \eta p_y$ pairing, where $\eta=\pm1$ denotes the chirality and $\langle \bm j,\bm k \rangle$ the nearest neighbors.
Here, $c_{\bm j}$ are fermionic annihilation operators for spinless fermions on site $\bm j$, satisfying the canonical anticommutation relations $\{c^\pdag_{\bm j}, c^\dag_{\bm k}\} = \delta_{\bm j,\bm k}$.
The parameter $t>0$ represents the hopping strength, $\mu$ is the chemical potential, and $\Delta_0>0$ determines the strength of the superconducting pairing.
The pairing term in the second line encodes the phase winding; the phase of the pairing amplitude depends on the spatial direction $\bm n = (\cos \phi, \sin\phi)^T$ as $e^{i \eta \phi}$, which is the defining property of a CPWSC.

The system resides in a topological phase as long as $\Delta_0 > 0$ and $|\mu| < 4t$, characterized by a Chern number of $C=-\eta$ of the lower Bogoliubov band ($\mu<0$) and $C=\eta$ for $\mu>0$; at $\mu =0$, the gap closes at $\bm{p}=(\pi/a,0), (0,\pi/a)$ with $a$ the lattice constant.
Consequently, a chiral Majorana edge state emerges at the boundary of the system.
To implement a minimal version of this CPWSC, the correlation length should be minimized to suppress finite-size effects.
This is achieved at the parameter choice $\mu = -2 \Delta_0$ and $t=\Delta_0 $, where the Bogoliubov spectrum assumes the form
\begin{equation}
  \epsilon_{\bm p} = 2 \Delta_0\sqrt{1+ 2[1-\cos (p_x a)] [1-\cos (p_y a)]}\,.
\end{equation}
The minimum gap occurs at $\bm p = 0$ with $\epsilon_0 = 2\Delta_0$, and the total bandwidth is $4\Delta_0$.

To clarify why this configuration constitutes a minimal model, we evaluate the edge-state decay length.
In 2D, the decay length depends on the orientation of the interface, characterized by its normal vector $\bm n$.
The decay length $\lambda$ of the chiral Majorana edge states orthogonal to the interface is determined by the root $\epsilon_{\kappa \bm n} = 0$ for a complex wave vector $\kappa\bm n$, so that $\lambda = 1/|\Im \kappa|$.
Remarkably, the decay length is zero (the wave function is confined to the outermost row) for interfaces along the $x$- and $y$-axes.
The least localized direction is the diagonal ($\phi=\pi/4$), for which we find $\lambda_\text{max} = \sqrt2\, a/\ln(3 + 2 \sqrt2)$.
Since the adjacent lattice site along the diagonal is at a distance of $\sqrt{2}\,a$, the probability amplitude drops to $\exp(- \sqrt{2}\,a/\lambda_\text{max}) = (3+2\sqrt2)^{-1} \approx 17\%$ from one site to the next.

Thus, if the interfaces are predominantly aligned with the crystalline axes, the edge modes of the CPWSC are maximally localized, justifying the designation `minimal' CPWSC.
Note that the experimental implementation discussed below is restricted to nearest-neighbor models, for which our choice of parameters is optimal; for a discussion of the underlying topological obstruction, see Refs.~\cite{thouless:84,chen:14}.
Including next-nearest-neighbor coupling could be used to further reduce the decay length~\cite{neupert:11}.

As a consequence, finite-size effects are weak already for small $L\times L$ lattices.
Along an interface parallel to a crystalline axis, the chiral edge mode disperses as $E_p = \pm2 \Delta_0 \sin (pa)$ with $|p| <\pi/(2a)$ the wave vector along the edge.
Since these states are almost perfectly confined to the edge and the leakage into the bulk is negligible, they extend over the full circumference $4La$.
The number of edge states in the gap is thus $N =  4La\, [\pi/(2a)]/(2\pi)=  L$ and their level spacing is $\delta = 2\pi \hbar v/(4La)$ with the edge velocity $v=2\Delta_0 a/\hbar$.
Numerical diagonalization confirms this estimate up to one level, see also below.

To illustrate how to realize this minimal model, we first describe a single link consisting of two quantum dots coupled via an Andreev dot.

\emph{Andreev link}--- 
The Hamiltonian of the Andreev dot is given by
\begin{equation}
	H_A =  \sum_{\sigma \in\{\uparrow, \downarrow\}} \xi_\sigma d^\dag_\sigma  d^\pdag_\sigma
	+ \Delta^* d_\downarrow d_\uparrow + \Delta d^\dag_\uparrow d^\dag_\downarrow \,.
\end{equation}
Here, we have introduced the fermionic annihilation operators $d_\sigma$ for spin $\sigma \in \{\uparrow,\downarrow\}$, that fulfill the canonical anticommutation relations.
The parameter $\xi_\sigma$ denotes the energy level of the Andreev dot (with respect to the chemical potential of the superconductor).
In the following, we assume $\xi=\xi_\uparrow=\xi_\downarrow>0$ for simplicity.
The parameter $\Delta = |\Delta| e^{i\varphi}$ describes the pairing strength due to the proximity-coupling to the superconductor.
Note that the effect of the magnetic field is typically smaller on the proximity-coupled quantum dots due to the Meissner screening and a reduced effective $g$-factor~\cite{vaitiekenas2018,dmytruk2018}.
As long as the Zeeman splitting on the Andreev dot is smaller than its excitation energy $\varepsilon= \sqrt{\xi^2 + |\Delta|^2}$, its ground state is the even-parity spin singlet.

The Andreev dot is coupled to two single-orbital quantum dots, with annihilation operators $c_{j\sigma}$ ($j\in\{1,2\}$), via the tunneling Hamiltonian $H_T=-\sum_j\bm d^\dag M_j\bm c_j+\text{H.c.}$, where $\bm d = (d_\uparrow\; d_\downarrow)^T$.
We first assume time-reversal-symmetric tunneling; orbital effects of the magnetic field are discussed below.
Time-reversal symmetry requires $\mathcal T^{-1}M_j\mathcal T=M_j$, where $\mathcal T=i\sigma_2K$ and $K$ denotes complex conjugation, giving the general form
\begin{equation}\label{eq:Mj}
	 M_j= t_j(\sigma_0+ i \alpha_j \bm e_j\cdot \bm \sigma )\,.
\end{equation}
The parameter $t_j>0$ describes the tunneling amplitude, $\bm e_j$ the effective direction of the spin-orbit interaction, and $\alpha_j\geq 0$ the strength of the spin-orbit interaction.
In fact, $\alpha_j$ is approximately given by the separation of the quantum dot $j$ to the Andreev dot divided by the spin-orbit length.

We are interested in a situation where the energy of the Andreev dot is sufficiently detuned from the energies in the normal dots; in particular, we need $\varepsilon \gg t_j$.
Integrating out the Andreev level in second-order perturbation theory (see End Matter), we obtain cotunneling with amplitude $\propto\xi t_k t_j/\varepsilon^2$ and crossed Andreev reflection with amplitude $\propto\Delta t_k t_j/\varepsilon^2$ between the dots, both weighted by the matrix $M_k^\dag M_j$.
Note that the prefactors depend only on the Andreev level, while the effective tunneling matrix $M_k^\dag M_j$ encodes the pairing symmetry.
Tuning the parameter $\xi$ of the Andreev level, the relative contribution of the cotunneling to the pairing can be changed.
In particular, by tuning gates, we can obtain a situation where the two effects are of the same strength.

In order to realize the minimal version of a CPWSC, we extend this result for a single link to a lattice with sites $\bm j$.
Moreover, effectively spinless fermions have to be generated by applying a sufficiently strong out-of-plane magnetic field $\bm B = B \hat {\bm z}$, $B>0$.
If the Zeeman energy is larger than the strength of both the cotunneling and the crossed Andreev reflection, we can project the model onto the single spin band $\uparrow$ (spin parallel to the field).
We assume that the spin-orbit fields lie in plane, $\bm e_{\bm j} \perp \hat {\bm z}$, and that the two parts of each link $\langle \bm j,\bm k\rangle$ are antialigned, $\bm e_{\bm j} = -\bm e_{\bm k}$.
Then, we obtain the expressions ($\bm e_- = \hat {\bm x} - i \hat {\bm y}$)
\begin{align}
  \label{eq:cot_car}
  H_\text{cot} &=-  
  \frac{\xi}{\varepsilon^2} \sum_{\langle \bm j,\bm k\rangle } t_{\bm k} t_{\bm j}(1-\alpha_{\bm k} \alpha_{\bm j} ) {c}_{\bm k}^\dag  {c}^\pdag_{\bm j} + \text{H.c.}\\
  H_\text{car} &= -i\frac{\Delta}{\varepsilon^2} \sum_{\langle \bm j,\bm k\rangle} t_{\bm k} t_{\bm j} (\alpha_{\bm k}  +\alpha_{\bm j} )(\bm e_{\bm j}\cdot \bm e_-)  \,c_{\bm k}^\dag   c_{\bm j}^\dag  + \text{H.c.}\,,\nonumber
\end{align}
for the cotunneling and the crossed Andreev reflection where $c_{\bm j} \equiv c_{\bm j\uparrow}$ denote the spinless fermionic operators; here, we have assumed that all the superconducting elements are connected such that there is a common phase.

\emph{Implementation in InAs and Ge}--- For this model (on a square lattice) to implement the minimal CPWSC, the direction of the spin-orbit interaction has to be properly aligned with the direction of the links, see Eq.~\eqref{eq:cp}.
In particular, for Rashba spin-orbit interaction in an InAs 2D electron gas, we have   $\bm e_j = (\sin \phi_j, -\cos\phi_j)^T$ where $(\cos \phi_j, \sin \phi_j)^T$ is the direction from the dot $j$ to the Andreev dot, while for heavy holes in Ge where the cubic Rashba interaction dominates~\cite{winkler2003,babkin2025}, $\bm e_j =  (\sin(3\phi_j), -\cos(3\phi_j))^T$.
Thus, $\bm e_j\cdot\bm e_-=i e^{-i\phi_j}$ (InAs) or $i e^{-3i\phi_j}$ (Ge); the opposite sign of the spin-orbit coupling reverses $\bm e_j$ and amounts to the shift $\varphi\mapsto\varphi+\pi$ of the superconducting phase.
We need to pick the directions $\phi_j$ of the links in such a way that the pairing on the links around a vertex has a phase difference of $\pm\pi/2$.
On the square lattice of Fig.~\ref{fig:setup}(b), with links along $\phi_j=0$ and $\pi/2$, the pairing phases are $1$ and $-i$ for InAs ($\eta=-1$), and $1$ and $e^{-3i\pi/2}=+i$ for Ge ($\eta=+1$).
Both materials thus realize the minimal CPWSC on a square lattice, albeit with opposite chirality.
For $\varphi=0$, the crossed Andreev reflection in Eq.~\eqref{eq:cot_car} then reproduces the pairing term of Eq.~\eqref{eq:cp} including its sign; a finite phase $\varphi$ corresponds to $\Delta_0\mapsto\Delta_0e^{i\varphi}$.
In order to fine-tune to the minimal CPWSC, we need to ensure that
\begin{equation*}
  \Delta_0 = \frac{\xi}{\varepsilon^2}  t_{\bm k} t_{\bm j}(1-\alpha_{\bm k} \alpha_{\bm j} ) \text{ and }  \Delta_0 = \frac{|\Delta|}{\varepsilon^2}  t_{\bm k} t_{\bm j} (\alpha_{\bm k}  +\alpha_{\bm j}),
\end{equation*}
on each link.
In tuning the chemical potential to $\mu =-2\Delta_0$, the renormalization of $\mu$ due to the diagonal terms $k=j$ in Eq.~\eqref{eq:h_eff} has to be taken into account.

A perpendicular field also induces orbital effects.
In the End Matter, we show that the $p$-wave pairing symmetry persists for integer multiples of $h/e$ per unit cell.
For $a=200\,\mathrm{nm}$, choosing $B=2h/(ea^2)\simeq0.21\,\mathrm{T}$ satisfies this condition.
Assuming $|g_\perp|\simeq10$ as representative for Ge and InAs~\cite{brosig2000,sommer2026}, this field yields a spin splitting $|g_\perp|\mu_B B\simeq120\,\mu\mathrm{eV}$.
For $t=\Delta_0\simeq20\,\mu\mathrm{eV}$ as in quantum dot Kitaev chains~\cite{tenhaaf2024}, this justifies the spinless approximation.
Moreover, hard induced gaps have been demonstrated for perpendicular fields of this magnitude, e.g., in planar Ge~\cite{fabris2026}, so that our proposal is within reach of current experiments.

\emph{Majorana modes in small arrays}---
\begin{figure}[tb]
  \centering
  \includegraphics[width=\linewidth]{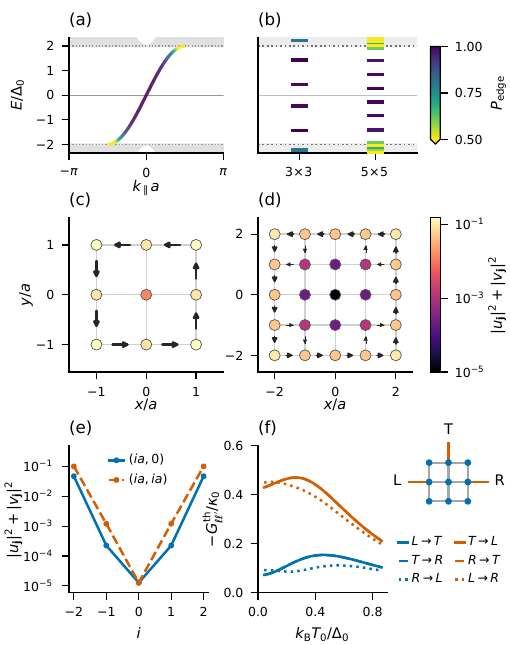}
  \caption{%
  Chiral edge states in a minimal CPWSC.
  (a) Dispersion of a straight edge with bulk bands indicated in gray.
  (b) Energy spectra of $3\times3$ and $5\times5$ arrays.
  Colors in (a),(b) give the boundary weight $P_{\mathrm{edge}}$: the probability on the outermost row in (a), and the boundary dots in (b).
  (c),(d) Quasiparticle-probability density of the lowest positive-energy eigenstate for the $3\times3$ and $5\times5$ arrays.
  Arrows show quasiparticle-probability currents with currents below $10^{-2}$ of the largest value omitted.
  (e) Horizontal and diagonal cuts of the density in (d).
  (f) Thermal conductances of a $3\times3$ array coupled to three normal leads shown in the inset.
  Leads are coupled strongly, with $t_c=2.56\Delta_0$.
  }\label{fig:chiral-edges}
\end{figure}
We numerically investigate finite square arrays described by Eq.~\eqref{eq:cp} at $t=\Delta_0$ and $\mu=-2\Delta_0$ with details of the calculations given in the End Matter.
Figure~\ref{fig:chiral-edges}(a) shows the chiral edge state dispersion at a straight boundary, while Fig.~\ref{fig:chiral-edges}(b) displays its discrete counterpart in small arrays.
Both $3\times3$ and $5\times5$ arrays show the approximately equally spaced edge state levels, as expected for an edge state with a linear dispersion relation.
The quasiparticle-probability densities and currents of the edge states exhibit strong localization near the boundary and circulation around the array [Figs.~\ref{fig:chiral-edges}(c),(d)].
The horizontal and diagonal cuts in Fig.~\ref{fig:chiral-edges}(e) show the strong suppression of the probability density toward the center by orders of magnitude even for these small arrays.
As expected, the horizontal cut shows a shorter decay length than the diagonal cut.

The chiral nature of the edge states is evident in thermal transport, even in very small dot arrays.
Figure~\ref{fig:chiral-edges}(f) shows the thermal conductance of a $3\times 3$ array with three normal leads.
With the quasiparticle current circulating counterclockwise [Fig.~\ref{fig:chiral-edges}(c)], thermal transport along this direction is considerably larger than in the opposite direction.

\begin{figure}[tb]
  \centering
  \includegraphics[width=\linewidth]{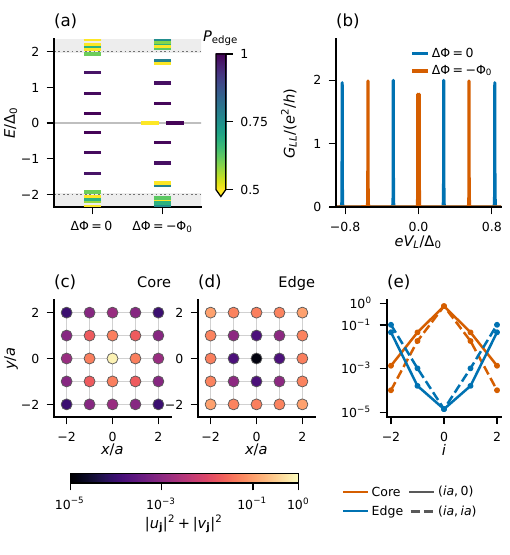}
  \caption{%
  Vortex and edge Majoranas in a $5\times 5$ array.
  (a) Spectra without and with an additional flux through the central superconducting square.
  Colors give the boundary weight $P_{\mathrm{edge}}$ of the wave functions, as in Fig.~\ref{fig:chiral-edges}(b).
  (b) Local conductance $G_{LL}$ of a single tunnel-coupled normal lead ($t_c =0.15\Delta_0$) connected to the middle of the left edge.
  (c),(d) Quasiparticle probabilities of the Majorana combinations that minimize (core) and maximize (edge) the boundary weight, represented in (a) by the left and right markers at zero mean energy, respectively.
  (e) Density cuts along the horizontal and diagonal axes of (c), (d).
  }\label{fig:vortex-majoranas}
\end{figure}

We next consider Majorana zero modes bound to vortices, the second signature of a CPWSC.
To this end, we introduce a flux $-\Phi_0=-\frac{h}{2e}$ through the central superconducting square of a $5\times5$ array.
This flux induces a pair of near-zero-energy states as shown in Fig.~\ref{fig:vortex-majoranas}(a) and shifts the overall spectrum accordingly.
Due to the finite size, these modes still have a small splitting $\pm E_0$, with $E_0\approx 0.00870\Delta_0$.
However, this splitting is much smaller than the level spacing, so that the zero-energy peak is clearly resolved in local tunneling measurements [Fig.~\ref{fig:vortex-majoranas}(b)].
For a $5\times 5$ array, the two Majorana components localize well at the vortex core and the outer boundary as seen in Fig.~\ref{fig:vortex-majoranas}(c)--(e).
Vortex-induced low-energy states already appear in a $3\times3$ array.
The overlap of the Majorana components in the core and at the edge, however, produces a splitting $E_0\approx 0.152\Delta_0$, an order of magnitude larger than in the $5\times5$ array.

\emph{Robustness to disorder}---
In a sweet-spot Kitaev chain~\cite{kitaev2001}, quasiparticles localize on single bonds, so that random dot energies of strength $W$ reduce the gap by an amount of order $W$.
In our 2D array, the lowest bulk states with $p_y=0$ ($p_x=0$) instead extend along a full column (row) of $L$ dots.
Projected onto the $p_y=0$ sector, the model reduces to a sweet-spot Kitaev chain along $x$ with dot energies $\overline{\delta\mu}_x = L^{-1}\sum_y\delta\mu_{(x,y)}$, whose fluctuations are reduced by a factor $\sqrt L$.
Hence, for the same number of dots, the 2D array is considerably more robust to disorder than the Kitaev chain (see End Matter for more information).

\emph{Conclusion}---
We have proposed a minimal 2D chiral $p$-wave superconductor, assembled from a square array of quantum dots coupled via proximitized Andreev dots.
We have shown that Rashba spin-orbit coupling in the tunnel barriers and a Zeeman field produce a spinless $p$-wave model, realizable on existing InAs and Ge quantum-dot arrays.
Tuned to its sweet spot, the array confines the edge states to the outermost row, so that chiral edge transport and a vortex-bound Majorana zero mode are already present in $3\times3$ and $5\times5$ arrays.
Moreover, the extended bulk modes of the 2D array average out disorder, so that the gap is more robust than in a Kitaev chain with the same number of dots.
Our results can be tested in tunneling and thermal-conductance measurements on present-day quantum-dot devices.
As a built-in control, gate tuning of the dot levels reverses the chirality when $\mu$ changes sign and yields a trivial phase for $|\mu|>4t$, where the chiral thermal conductance and the vortex Majorana zero mode disappear.
Braiding vortex Majoranas by moving them around each other, which would implement braiding of Ising anyons, remains an interesting question for future research.

\emph{Acknowledgments}--- We have benefited from discussions with B.\ van Heck. FH acknowledges support by the Deutsche Forschungsgemeinschaft (DFG) under Germany's Excellence Strategy (Cluster of Excellence Matter and Light for Quantum Computing ML4Q, EXC 2004/1 -- 390534769). MW acknowledges OpenAI for providing access to the ChatGPT for Academic Researchers program.
  AI tools (Claude Opus 5.5 and GPT-6 Astra), closely steered by the authors, assisted with calculations, numerics, and text polishing; the authors take full responsibility for the content.

\emph{Data availability}---
The code and data underlying the figures are available at Zenodo \cite{zenodo}.

\clearpage
\section*{End Matter}

\newcounter{emsec}
\counterwithin*{equation}{emsec} 
\newcommand{\emsection}[1]{\stepcounter{emsec}\emph{#1}---}
\renewcommand{\theequation}{\Alph{emsec}\arabic{equation}}
\renewcommand{\theHequation}{endmatter.\Alph{emsec}.\arabic{equation}}

\emsection{Andreev link}
The Andreev dot of the main text can be diagonalized by introducing the Bogoliubov operators $\bm \beta = u \bm d + v i \sigma_2 \bm d^\dag$, where $\bm d = (d_\uparrow\; d_\downarrow)^T$, $u=\sqrt{(1+\xi/\varepsilon)/2}$, and $v =e^{i\varphi}\sqrt{(1-\xi/\varepsilon)/2}$.
The Hamiltonian of the Andreev dot assumes the form $ H_A =  \varepsilon \sum_\sigma \beta^\dag_\sigma \beta^\pdag_\sigma + \xi - \varepsilon$ in the new basis.
Expressed in the Bogoliubov operators, the tunneling Hamiltonian becomes
\begin{align}\label{eq:tunneling-general}
 H_T&=-\sum_j\bm d^\dag M_j\bm c_j+\text{H.c.}\nonumber\\
 &=-\sum_j\bm\beta^\dag\bigl[uM_j\bm c_j-v i\sigma_2M_j^*\bm c_j^\dag\bigr]+\text{H.c.}\,.
\end{align}
For time-reversal-symmetric tunneling, Eq.~\eqref{eq:Mj}, we have $i\sigma_2M_j^*=M_ji\sigma_2$, and Eq.~\eqref{eq:tunneling-general} simplifies to $H_T=-\sum_j\bm\beta^\dag M_j[u\bm c_j-v i\sigma_2\bm c_j^\dag]+\text{H.c.}$.
For $\varepsilon \gg t_j$, we integrate out the Andreev level in second-order perturbation theory and arrive at an effective Hamiltonian for the normal dots of the form
\begin{equation}\label{eq:h_eff}
	H_\text{eff} =   -\frac{1}{\varepsilon}\sum_{j,k} (u \bm c_k^\dag +v^* {\bm c}_k i \sigma_2 ) M_{kj} [u \bm c_j -v i \sigma_2 {\bm c}^\dag_j]\,.
\end{equation}
The Hamiltonian depends on the matrix $M_{kj}= M_{k}^\dag M^\pdag_j= t_k t_j( m^{(0)}_{kj} \sigma_0 + i \bm m_{kj} \cdot {\bm \sigma}) $ with
\begin{align}
  m^{(0)}_{kj} &= 1 + \alpha_k \alpha_j \bm e_k \cdot \bm e_j\,, \nonumber \\
  \bm m_{kj} &= \alpha_j \bm e_j  -\alpha_k \bm e_k
  +\alpha_k \alpha_j \bm e_k \times \bm e_j\,,
\end{align}
which encodes the transfer of a particle from dot $j$ to dot $k$; note that $m^{(0)}$ is symmetric under exchange of $k$ and $j$ while $\bm m$ is antisymmetric.
The result, Eq.~\eqref{eq:h_eff}, remains valid for an arbitrary number of normal dots coupled to a single Andreev dot.
In particular, the `diagonal' terms with $k=j$ describe the renormalization of the dot parameters due to the exchange with the Andreev dot.

We can rewrite the effective Hamiltonian as $H_\text{eff} = H_\text{cot} + H_\text{car}$, separating the cotunneling contribution from the crossed Andreev processes.
The cotunneling is given by
\begin{align}
H_\text{cot} &=  -\frac{1}{\varepsilon}
\sum_{j,k} {\bm c}^\dag_k (u^2-|v|^2)M_{kj} {\bm c}_j \\
&= -\frac{\xi}{\varepsilon^2} \sum_{j,k} t_k t_j {\bm c}_{k}^\dag
 (m^{(0)}_{kj}\sigma_0+i\bm m_{kj}\cdot\bm\sigma){\bm c}_{j}\,.\nonumber
\end{align}
More importantly, the crossed Andreev reflection assumes the form $H_\text{car} = \sum_{j,k} \bm c_k^\dag \Delta_{kj} {\bm c}_j^\dag + \text{H.c.}$ with the pairing term
\begin{equation}
  \Delta_{kj} = \frac{\Delta}{2 \varepsilon^2} t_k t_j ( m^{(0)}_{kj} \sigma_0 + i{\bm m}_{kj}\cdot \bm\sigma) i \sigma_2\,.
\end{equation}
Projecting onto the spin-$\uparrow$ band for in-plane and antialigned spin-orbit fields, $\bm e_{\bm j}=-\bm e_{\bm k}\perp\hat{\bm z}$, yields Eq.~\eqref{eq:cot_car} of the main text.

\emsection{Orbital effect of a magnetic field}
We include the orbital effect of a magnetic field by multiplying each tunneling matrix by a Peierls phase factor,
\begin{equation}\label{eq:orbital-peierls}
 M_{\bm j}\longmapsto P_{\bm j}M_{\bm j},\qquad
 P_{\bm j}=\exp\!\left[-\frac{ie}{\hbar}\int_{\bm j}^M\bm A\cdot d\bm\ell\right],
\end{equation}
where the electron charge is $-e$, $e>0$, $\bm A$ is the magnetic vector potential, $\bm j$ denotes the position of the normal dot, and $M$ the position of the Andreev dot.
Substituting into Eq.~\eqref{eq:tunneling-general}, and using time-reversal symmetry only for the bare tunneling $M_{\bm j}$, gives
\begin{align}
   H_T&=-\sum_{\bm j}\bm\beta^\dag M_{\bm j}\bigl[uP_{\bm j}\bm c_{\bm j}-vP_{\bm j}^*i\sigma_2\bm c_{\bm j}^\dag\bigr]+\text{H.c.}\,.
\end{align}
Hence, on a bond from $\bm j$ to $\bm k$, cotunneling processes acquire a phase factor $P_{\bm k}^*P_{\bm j}$, whereas pair creation acquires $P_{\bm k}^*P_{\bm j}^*$ in addition to the local superconducting phase.
Incorporating these phases into Eq.~\eqref{eq:cot_car} results in the $p$-wave Hamiltonian, Eq.~\eqref{eq:cp}, with $t\mapsto te^{i\lambda_{\bm k\bm j}}$ and $\Delta_0\mapsto\Delta_0e^{i\theta_{\bm k\bm j}}$, where
\begin{align}
 \lambda_{\bm k\bm j}&=-\frac{e}{\hbar}\int_{\bm j}^{\bm k}\bm A\cdot d\bm\ell,
 \label{eq:orbital-hopping}\\
 \theta_{\bm k\bm j}&=\varphi_M +\frac{e}{\hbar}\left(\int_{\bm j}^M\bm A\cdot d\bm\ell-\int_M^{\bm k}\bm A\cdot d\bm\ell\right);
 \label{eq:orbital-pairing}
\end{align}
here, $\varphi_M$ is the superconducting phase of the Andreev dot.

\emsection{Uniform background field}
We now consider a square lattice of dots with interlaced, thin superconducting wires, as in Fig.~\ref{fig:setup}(b).
The normal dots are located at $a(n_x,n_y)$, with lattice spacing $a$ and integers $n_x,n_y$.
The Andreev dots sit at $a(n_x+\tfrac12,n_y)$ and $a(n_x,n_y+\tfrac12)$, connected by a network of superconducting wires along $x=(n_x+\tfrac12)a$ and $y=(n_y+\tfrac12)a$.

Superconducting wires that are thin compared to the London penetration depth do not screen the magnetic field, and we thus assume a uniform out-of-plane field $B_z = m \frac{h}{e a^2}$ where $m$ counts the number of electron flux quanta $\frac{h}{e}$ through each unit cell.

In Landau gauge, the vector potential is $\bm A = B_z x \hat{\bm y}$, and Eq.~\eqref{eq:orbital-hopping} gives $\lambda_x=0$ and $\lambda_y=-2\pi m n_x$.
Hence, for integer $m$, the hopping term is unchanged.
At the same time, the Peierls phase factors in Eq.~\eqref{eq:orbital-pairing} cancel for any uniform field, giving $\theta_{\bm k\bm j}=\varphi_M$.

To determine the phases $\varphi_M$, we note that for integer $m$, the lowest energy configuration has zero supercurrent along the wires, and thus $\partial_\parallel\varphi=-\frac{2e}{\hbar}A_\parallel$, which is compatible with a single-valued order parameter as the phase accumulated around any superconducting square is $-4\pi m$.
Choosing the phase $0$ on every horizontal wire, we find $\varphi_M=0$ for vertical dot links and $\varphi_M=-\pi m(2n_x+1)$ modulo $2\pi$ for horizontal dot links, where the Andreev dot lies at $y=n_y a$ on a vertical wire; see the Supplemental Material~\cite{supplemental} for details.
Thus, horizontal links acquire the pairing phase $e^{i\theta_{\bm k\bm j}}=(-1)^m$, i.e., $c_{\bm j+\hat{\bm x}}^\dag\mapsto(-1)^m c_{\bm j+\hat{\bm x}}^\dag$ in the pairing term of Eq.~\eqref{eq:cp}.
Hence, even $m$ gives $p_x+i\eta p_y$ pairing, while odd $m$ reverses it to $p_x-i\eta p_y$.

\emsection{Square vortex}
To add a vortex, we add an extra flux $s\Phi_0$, with $s=\pm1$ and $\Phi_0=h/(2e)$, uniformly over the central superconducting square $|x|,|y|<a/2$.
Using a fourfold-symmetric gauge for the vortex field, we find that the vortex gives an additional phase to the pairing term as
\begin{equation}\label{eq:square-vortex-factor}
e^{i\theta_{\bm k\bm j}^v}
 =\frac{z_{\bm j}^{(s)}+z_{\bm k}^{(s)}}{|z_{\bm j}^{(s)}+z_{\bm k}^{(s)}|},
\end{equation}
where we define
\begin{equation}\label{eq:square-vortex-z}
 z_{\bm j}^{(s)}=
 \frac{\sgn(x_{\bm j})-is\sgn(y_{\bm j})}
 {\sqrt{\sgn^2(x_{\bm j})+\sgn^2(y_{\bm j})}},
 \qquad z_{\bm 0}^{(s)}=0.
\end{equation}
with the convention $\sgn(0)=0$, and $z_{\bm 0}^{(s)}=0$
for the central dot $(x_{\bm j},y_{\bm j})=(0,0)$.

Away from the coordinate axes, the pairing phase is constant within each quadrant, giving a discrete ``square'' vortex.
A smooth polar-angle profile is not required to bind a Majorana mode: discrete vortices can also support such states~\cite{fu2008}.

The vortex field also enters the hopping phases, Eq.~\eqref{eq:orbital-hopping}; see the Supplemental Material~\cite{supplemental} for details.

\emsection{Numerical simulations}
We implement the $p$-wave Hamiltonian, Eq.~\eqref{eq:cp}, in Bogoliubov--de Gennes (BdG) form using Kwant~\cite{groth2014}, including the pairing and orbital phases derived above.
The BdG Hamiltonian is
\begin{align}\label{eq:numerical-bdg}
  \mathcal H_{\mathrm{BdG}}&=-\mu\tau_z+\sum_{\langle\bm j,\bm k\rangle}\left(
 |\bm k\rangle\langle\bm j|\otimes\mathcal H_{\bm k\bm j}+\mathrm{H.c.}\right).
\end{align}
Here $|\bm j\rangle$ labels the dot, and $\tau_{x,y,z}$ act in the local Nambu basis $(c_{\bm j},c_{\bm j}^\dag)$.
The bond blocks are
\begin{align}\label{eq:numerical-bdg-bond}
  \mathcal H_{\bm k\bm j}&=-t e^{i\lambda_{\bm k\bm j}\tau_z}\tau_z+i\Delta_0 e^{i\theta_{\bm k\bm j}\tau_z}
 \begin{cases}
 \tau_y,&\bm k=\bm j+\hat{\bm x},\\
 -\tau_x,&\bm k=\bm j+\hat{\bm y}.
 \end{cases}
\end{align}
Unless stated otherwise, we use  $\eta=-1$, $t=\Delta_0$, and $\mu=-2\Delta_0$, and measure lengths and energies in units of $a$ and $\Delta_0$.

For isolated $L\times L$ lattices, we obtain spectra and wave functions by numerical diagonalization.
For a normalized eigenstate $\psi_{\bm j}=(u_{\bm j},v_{\bm j})^T$, the quasiparticle probability on dot $\bm j$ is $|u_{\bm j}|^2+|v_{\bm j}|^2$.
The quasiparticle current from $\bm j$ to $\bm k$ is $J_{\bm j\to\bm k}=(2/\hbar)\Im(\psi_{\bm k}^\dag\mathcal H_{\bm k\bm j}\psi_{\bm j})$, including both normal and anomalous bond terms.
To separate vortex core and edge Majorana wave functions, we form the two Majorana combinations within the lowest particle-hole symmetric pair that minimize and maximize the weight on the outermost dots, see the Supplemental Material~\cite{supplemental}.

For thermal transport, we attach three semi-infinite spinless normal chains to the left, top, and right side midpoints.
For tunneling spectroscopy, only the left chain is attached.
The leads have zero onsite energy, hopping $t_\ell=1.5\Delta_0$, and contact hopping $t_c$ to the island.
We use Kwant to compute the scattering matrix $S(E)$ of the system, and from it the scattering probabilities $T_{\ell\ell'}^{\alpha\beta}(E)$ from lead $\ell'$ to lead $\ell$, with $\ell,\ell'\in\{L,T,R\}$ and outgoing and incoming quasiparticle types $\alpha,\beta\in\{e,h\}$, respectively.
We obtain the charge and thermal conductances from these probabilities using the scattering formalism reviewed in Ref.~\cite{waintal2024}; the explicit expressions are given in the Supplemental Material~\cite{supplemental}.
We report thermal conductances in units of $\kappa_0=\pi^2k_B^2T/(3h)$.

\begin{figure}[tb]
  \centering
  \includegraphics[width=\linewidth]{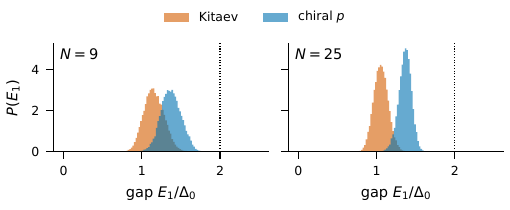}
  \caption{%
    Distribution of the lowest positive BdG eigenvalue $E_1$ (the bulk gap) for the minimal CPWSC on an $L\times L$ torus (2D) and a sweet-spot Kitaev ring (1D) with the same number $N=L^2$ of dots, at $N=9$ (left) and $N=25$ (right), for disorder strength $W=3\Delta_0$.
    Each histogram contains $2\times10^4$ disorder realizations.
    The dotted line indicates the clean gap $2\Delta_0$.
  }\label{fig:disorder}
\end{figure}

\emsection{Disorder}
To benchmark disorder robustness, we add random dot energies $\delta\mu_{\bm j}$, drawn uniformly from $[-W/2,W/2]$, to an $L\times L$ torus of the minimal CPWSC and to a sweet-spot Kitaev ring ($t=\Delta_0$, $\mu=0$).
Both systems have the same number $N=L^2$ of dots and share the clean gap $2\Delta_0$.
Figure~\ref{fig:disorder} shows the resulting distributions of the lowest positive BdG eigenvalue $E_1$.
At the same number of dots, disorder reduces the gap of the Kitaev ring considerably more than that of the 2D array, and the difference grows with $N$.
The mean gap of the 2D array is essentially independent of $N$ ($\approx1.37\Delta_0$ for $N=9$ and $\approx1.36\Delta_0$ for $N=25$) while its distribution narrows, whereas the mean gap of the Kitaev ring decreases from $\approx1.16\Delta_0$ to $\approx1.06\Delta_0$.
This reflects the mechanism discussed in the main text: the lowest bulk states of the 2D array extend over a full row or column of $L$ dots, which averages the random dot energies and reduces their fluctuations by a factor $\sqrt{L}$, whereas the quasiparticles of the Kitaev ring are localized on single bonds.

\clearpage

\begin{thebibliography}{10}
\makeatletter\auto@bib@empty\let\auto@bib@innerbib\@empty\makeatother

\bibitem{read2000}
N. Read and D. Green,
 Paired states of fermions in two dimensions with breaking of parity and
  time-reversal symmetries and the fractional quantum Hall effect,
 \href{http://dx.doi.org/10.1103/PhysRevB.61.10267}{%
 Phys. Rev. B {\bf 61}, 10267 (2000)}.

\bibitem{alicea2012}
J. Alicea,
 New directions in the pursuit of Majorana fermions in solid state systems,
 \href{http://dx.doi.org/10.1088/0034-4885/75/7/076501}{%
 Rep. Prog. Phys. {\bf 75}, 076501 (2012)}.

\bibitem{ivanov2001}
D. A. Ivanov,
 Non-{A}belian statistics of half-quantum vortices in $p$-wave superconductors,
 \href{http://dx.doi.org/10.1103/PhysRevLett.86.268}{%
 Phys. Rev. Lett. {\bf 86}, 268 (2001)}.

\bibitem{nayak2008}
C. Nayak, S. H. Simon, A. Stern, M. Freedman, and S. Das~Sarma,
 Non-{A}belian anyons and topological quantum computation,
 \href{http://dx.doi.org/10.1103/RevModPhys.80.1083}{%
 Rev. Mod. Phys. {\bf 80}, 1083 (2008)}.

\bibitem{moore1991}
G. Moore and N. Read,
 Nonabelions in the fractional quantum Hall effect,
 \href{http://dx.doi.org/10.1016/0550-3213(91)90407-O}{%
 Nucl. Phys. B {\bf 360}, 362 (1991)}.

\bibitem{banerjee2018}
M. Banerjee, M. Heiblum, V. Umansky, D. E. Feldman, Y. Oreg, and A. Stern,
 Observation of half-integer thermal Hall conductance,
 \href{http://dx.doi.org/10.1038/s41586-018-0184-1}{%
 Nature {\bf 559}, 205 (2018)}.

\bibitem{dutta2022}
B. Dutta, W. Yang, R. Melcer, H. K. Kundu, M. Heiblum, V. Umansky, Y. Oreg, A.
  Stern, and D. Mross,
 Distinguishing between non-{A}belian topological orders in a quantum Hall
  system,
 \href{http://dx.doi.org/10.1126/science.abg6116}{%
 Science {\bf 375}, 193 (2022)}.

\bibitem{kallin2016}
C. Kallin and J. Berlinsky,
 Chiral superconductors,
 \href{http://dx.doi.org/10.1088/0034-4885/79/5/054502}{%
 Rep. Prog. Phys. {\bf 79}, 054502 (2016)}.

\bibitem{rice1995}
T. M. Rice and M. Sigrist,
 Sr$_2${R}u{O}$_4$: an electronic analogue of $^{3}${H}e?,
 \href{http://dx.doi.org/10.1088/0953-8984/7/47/002}{%
 J. Phys.: Condens. Matter {\bf 7}, L643 (1995)}.

\bibitem{maeno2024}
Y. Maeno, A. Ikeda, and G. Mattoni,
 Thirty years of puzzling superconductivity in Sr$_2${R}u{O}$_4$,
 \href{http://dx.doi.org/10.1038/s41567-024-02656-0}{%
 Nat. Phys. {\bf 20}, 1712 (2024)}.

\bibitem{jiao2020}
L. Jiao, S. Howard, S. Ran, Z. Wang, J. O. Rodriguez, M. Sigrist, Z. Wang, N.
  P. Butch, and V. Madhavan,
 Chiral superconductivity in heavy-fermion metal {UT}e$_2$,
 \href{http://dx.doi.org/10.1038/s41586-020-2122-2}{%
 Nature {\bf 579}, 523 (2020)}.

\bibitem{aoki2022}
D. Aoki, J.-P. Brison, J. Flouquet, K. Ishida, G. Knebel, Y. Tokunaga, and Y.
  Yanase,
 Unconventional superconductivity in {UT}e$_2$,
 \href{http://dx.doi.org/10.1088/1361-648X/ac5863}{%
 J. Phys.: Condens. Matter {\bf 34}, 243002 (2022)}.

\bibitem{han2025}
T. Han {\em et~al.\/},
 Signatures of chiral superconductivity in rhombohedral graphene,
 \href{http://dx.doi.org/10.1038/s41586-025-09169-7}{%
 Nature {\bf 643}, 654 (2025)}.

\bibitem{fu2008}
L. Fu and C. L. Kane,
 Superconducting proximity effect and Majorana fermions at the surface of a
  topological insulator,
 \href{http://dx.doi.org/10.1103/PhysRevLett.100.096407}{%
 Phys. Rev. Lett. {\bf 100}, 096407 (2008)}.

\bibitem{qi2010}
X.-L. Qi, T. L. Hughes, and S.-C. Zhang,
 Chiral topological superconductor from the quantum Hall state,
 \href{http://dx.doi.org/10.1103/PhysRevB.82.184516}{%
 Phys. Rev. B {\bf 82}, 184516 (2010)}.

\bibitem{lutchyn2010}
R. M. Lutchyn, J. D. Sau, and S. Das~Sarma,
 Majorana fermions and a topological phase transition in
  semiconductor-superconductor heterostructures,
 \href{http://dx.doi.org/10.1103/PhysRevLett.105.077001}{%
 Phys. Rev. Lett. {\bf 105}, 077001 (2010)}.

\bibitem{oreg2010}
Y. Oreg, G. Refael, and F. von Oppen,
 Helical liquids and Majorana bound states in quantum wires,
 \href{http://dx.doi.org/10.1103/PhysRevLett.105.177002}{%
 Phys. Rev. Lett. {\bf 105}, 177002 (2010)}.

\bibitem{coppini2026}
V. Coppini and B. van Heck,
 Chiral topological phases of proximitized planar germanium,
 \href{http://dx.doi.org/10.48550/arXiv.2611.17579}{%
 arXiv:2611.17579 (2026)}.

\bibitem{flensberg2021}
K. Flensberg, F. von Oppen, and A. Stern,
 Engineered platforms for topological superconductivity and Majorana zero
  modes,
 \href{http://dx.doi.org/10.1038/s41578-021-00336-6}{%
 Nat. Rev. Mater. {\bf 6}, 944 (2021)}.

\bibitem{sau2012}
J. D. Sau and S. Das~Sarma,
 Realizing a robust practical Majorana chain in a quantum-dot-superconductor
  linear array,
 \href{http://dx.doi.org/10.1038/ncomms1966}{%
 Nat. Commun. {\bf 3} (1), 964 (2012)}.

\bibitem{leijnse2012}
M. Leijnse and K. Flensberg,
 Parity qubits and poor man's Majorana bound states in double quantum dots,
 \href{http://dx.doi.org/10.1103/PhysRevB.86.134528}{%
 Phys. Rev. B {\bf 86}, 134528 (2012)}.

\bibitem{fulga2013}
I. C. Fulga, A. Haim, A. R. Akhmerov, and Y. Oreg,
 Adaptive tuning of Majorana fermions in a quantum dot chain,
 \href{http://dx.doi.org/10.1088/1367-2630/15/4/045020}{%
 New J. Phys. {\bf 15} (4), 045020 (2013)}.

\bibitem{liu2022}
C.-X. Liu, G. Wang, T. Dvir, and M. Wimmer,
 Tunable superconducting coupling of quantum dots via Andreev bound states in
  semiconductor-superconductor nanowires,
 \href{http://dx.doi.org/10.1103/PhysRevLett.129.267701}{%
 Phys. Rev. Lett. {\bf 129}, 267701 (2022)}.

\bibitem{dvir2023}
T. Dvir, G. Wang, N. van Loo, C.-X. Liu, G. P. Mazur, A. Bordin, S. L. D. ten
  Haaf, J.-Y. Wang, D. van Driel, F. Zatelli, X. Li, F. K. Malinowski, S.
  Gazibegovic, G. Badawy, E. P. A. M. Bakkers, M. Wimmer, and L. P.
  Kouwenhoven,
 Realization of a minimal Kitaev chain in coupled quantum dots,
 \href{http://dx.doi.org/10.1038/s41586-022-05585-1}{%
 Nature {\bf 614} (7948), 445 (2023)}.

\bibitem{tenhaaf2024}
S. L. D. ten Haaf, Q. Wang, A. M. Bozkurt, C.-X. Liu, I. Kulesh, P. Kim, D.
  Xiao, C. Thomas, M. J. Manfra, T. Dvir, M. Wimmer, and S. Goswami,
 A two-site Kitaev chain in a two-dimensional electron gas,
 \href{http://dx.doi.org/10.1038/s41586-024-07434-9}{%
 Nature {\bf 630} (8016), 329 (2024)}.

\bibitem{bordin2025}
A. Bordin, C.-X. Liu, T. Dvir, F. Zatelli, S. L. D. ten Haaf, D. van Driel, G.
  Wang, N. van Loo, Y. Zhang, J. C. Wolff, T. Van~Caekenberghe, G. Badawy, S.
  Gazibegovic, E. P. A. M. Bakkers, M. Wimmer, L. P. Kouwenhoven, and G. P.
  Mazur,
 Enhanced Majorana stability in a three-site Kitaev chain,
 \href{http://dx.doi.org/10.1038/s41565-025-01894-4}{%
 Nat. Nanotechnol. {\bf 20} (6), 726 (2025)}.

\bibitem{tenhaaf2025}
S. L. D. ten Haaf, Y. Zhang, Q. Wang, A. Bordin, C.-X. Liu, I. Kulesh, V. P. M.
  Sietses, C. G. Prosko, D. Xiao, C. Thomas, M. J. Manfra, M. Wimmer, and S.
  Goswami,
 Observation of edge and bulk states in a three-site Kitaev chain,
 \href{http://dx.doi.org/10.1038/s41586-025-08892-5}{%
 Nature {\bf 641} (8064), 890 (2025)}.

\bibitem{Note1}
`Minimal' refers to tight-binding models with nearest-neighbor couplings; only
  such local couplings define a 2D geometry and thus an edge that can carry a
  chiral Majorana mode. Like the minimal Kitaev chain, the array can be scaled
  up to larger sizes.

\bibitem{thouless:84}
D. J. Thouless,
 Wannier functions for magnetic sub-bands,
 \href{http://dx.doi.org/10.1088/0022-3719/17/12/003}{%
 J. Phys. C: Solid State Phys. {\bf 17} (12), L325 (1984)}.

\bibitem{chen:14}
L. Chen, T. Mazaheri, A. Seidel, and X. Tang,
 The impossibility of exactly flat non-trivial Chern bands in strictly local
  periodic tight binding models,
 \href{http://dx.doi.org/10.1088/1751-8113/47/15/152001}{%
 J. Phys. A: Math. Theor. {\bf 47} (15), 152001 (2014)}.

\bibitem{neupert:11}
T. Neupert, L. Santos, C. Chamon, and C. Mudry,
 Fractional quantum Hall states at zero magnetic field,
 \href{http://dx.doi.org/10.1103/PhysRevLett.106.236804}{%
 Phys. Rev. Lett. {\bf 106}, 236804 (2011)}.

\bibitem{vaitiekenas2018}
S. Vaitiek\.{e}nas, M.-T. Deng, J. Nyg{\aa}rd, P. Krogstrup, and C. M. Marcus,
 Effective $g$ factor of subgap states in hybrid nanowires,
 \href{http://dx.doi.org/10.1103/PhysRevLett.121.037703}{%
 Phys. Rev. Lett. {\bf 121}, 037703 (2018)}.

\bibitem{dmytruk2018}
O. Dmytruk, D. Chevallier, D. Loss, and J. Klinovaja,
 Renormalization of the quantum dot $g$-factor in superconducting Rashba
  nanowires,
 \href{http://dx.doi.org/10.1103/PhysRevB.98.165403}{%
 Phys. Rev. B {\bf 98}, 165403 (2018)}.

\bibitem{winkler2003}
R. Winkler,
 {\em Spin--Orbit Coupling Effects in Two-Dimensional Electron and Hole
  Systems\/}, vol. 191 of {\em Springer Tracts in Modern Physics\/}
 (Springer, Berlin, Heidelberg, 2003).

\bibitem{babkin2025}
S. S. Babkin, B. Joecker, K. Flensberg, M. Serbyn, and J. Danon,
 Superconducting proximity effect in two-dimensional hole gases,
 \href{http://dx.doi.org/10.1103/k4jh-pnxy}{%
 Phys. Rev. B {\bf 111}, 214518 (2025)}.

\bibitem{brosig2000}
S. Brosig, K. Ensslin, A. G. Jansen, C. Nguyen, B. Brar, M. Thomas, and H.
  Kroemer,
 {InAs-AlSb} quantum wells in tilted magnetic fields,
 \href{http://dx.doi.org/10.1103/PhysRevB.61.13045}{%
 Phys. Rev. B {\bf 61}, 13045 (2000)}.

\bibitem{sommer2026}
L. Sommer, I. Seidler, F. J. Schupp, S. Paredes, N. W. Hendrickx, L. Massai, K.
  Tsoukalas, A. Orekhov, E. G. Kelly, S. W. Bedell, G. Salis, M. Mergenthaler,
  P. Harvey-Collard, A. Fuhrer, and T. Ihn,
 Disentangling orbital and confinement contributions to g-factor in {Ge/SiGe}
  hole quantum dots,
 \href{http://dx.doi.org/10.1021/acs.nanolett.6c00999}{%
 Nano Lett. {\bf 26} (25), 8117 (2026)}.

\bibitem{fabris2026}
G. Fabris, P. Falthansl-Scheinecker, D. Shah, D. M. Pino, M. Borovkov, A.
  Bubis, K. Roux, D. Sokolova, A. A. Juanes, T. Costanzo, I. Taha, A. Genç, J.
  Arbiol, S. Calcaterra, A. D. C. Oliveira, D. Chrastina, G. Isella, R. S.
  Souto, M. J. Calderon, R. Aguado, J. C. Abadillo-Uriel, and G. Katsaros,
 Granular aluminum induced superconductivity in germanium for hole spin-based
  hybrid devices,
 \href{http://dx.doi.org/10.48550/arXiv.2602.21364}{%
 arXiv:2602.21364 (2026)}.

\bibitem{kitaev2001}
A. Y. Kitaev,
 Unpaired Majorana fermions in quantum wires,
 \href{http://dx.doi.org/10.1070/1063-7869/44/10S/S29}{%
 Phys.-Usp. {\bf 44}, 131 (2001)}.

\bibitem{zenodo}
F. Hassler and M. Wimmer,
 Code and data for ``{M}inimal chiral $p$-wave superconductor'',
 \href{http://dx.doi.org/10.5281/zenodo.24372227}{%
 Zenodo (2026)}.

\bibitem{supplemental}
See Supplemental Material for the derivation of the orbital phases and the
  square-vortex pairing profile.

\bibitem{groth2014}
C. W. Groth, M. Wimmer, A. R. Akhmerov, and X. Waintal,
 {Kwant}: a software package for quantum transport,
 \href{http://dx.doi.org/10.1088/1367-2630/16/6/063065}{%
 New J. Phys. {\bf 16}, 063065 (2014)}.

\bibitem{waintal2024}
X. Waintal, M. Wimmer, A. Akhmerov, C. Groth, B. K. Nikoli\'{c}, M. Istas, T.
  \"O. Rosdahl, and D. Varjas,
 Computational quantum transport: a scattering approach perspective,
 \href{http://dx.doi.org/10.48550/arXiv.2407.16257}{%
 arXiv:2407.16257 (2024)}.

\end{thebibliography}
\end{document}


\title{Supplemental Material: Minimal chiral \emph{p}-wave superconductor}
\maketitle
\renewcommand{\theequation}{S\arabic{equation}}
\section{Orbital phases from the Andreev dot}
\label{sec:sm-phases}

We start from the Andreev dot of the main text. For two normal dots
coupled through an Andreev dot at $M$, its Hamiltonian and the tunneling
Hamiltonian are
\begin{align}
 H_A&=\xi\sum_\sigma d_\sigma^\dagger d_\sigma
 +\Delta^*d_\downarrow d_\uparrow
 +\Delta d_\uparrow^\dagger d_\downarrow^\dagger,
 \qquad \Delta=|\Delta|e^{i\varphi_M},\label{eq:sm-abs}\\
 H_T&=-\sum_{j=1,2}\bm d^\dagger M_j\bm c_j+\mathrm{H.c.},
 \qquad M_j=t_j(\sigma_0+i\alpha_j\bm e_j\cdot\bm\sigma).
 \label{eq:sm-tunneling}
\end{align}
Here $\bm d=(d_\uparrow,d_\downarrow)^T$ and
$\bm c_j=(c_{j\uparrow},c_{j\downarrow})^T$ are spinors, and $M_j$
describes tunneling from normal dot $j$ into the Andreev dot. As in the
main text, we eliminate the Andreev dot perturbatively for
$\varepsilon=\sqrt{\xi^2+|\Delta|^2}\gg t_j$, and project the normal dots
onto the retained spin band. We derive the orbital phases of the resulting
cotunneling and crossed-Andreev terms while keeping their magnitudes
and spin structure fixed.

An electron has charge $-e$, with $e>0$. Its tunneling amplitude from
$j$ to $M$ therefore acquires the Peierls factor
\begin{equation}
 M_j\longmapsto
 e^{-\frac{ie}{\hbar}\int_j^M\bm A\cdot d\bm\ell}M_j.
 \label{eq:sm-inward}
\end{equation}
The integral follows the tunnel path toward the Andreev dot. The reverse
amplitude, multiplying $\bm c_j^\dagger\bm d$, is its Hermitian
conjugate and carries the opposite phase.

Co-tunneling from $j$ to $k$ involves an inward amplitude at $j$ and an
outward amplitude at $k$. Consequently, the matrix product
$M_k^\dagger M_j$ in the effective Hamiltonian acquires the factor
\begin{equation}
 e^{+\frac{ie}{\hbar}\int_k^M\bm A\cdot d\bm\ell}
 e^{-\frac{ie}{\hbar}\int_j^M\bm A\cdot d\bm\ell}
 =e^{-\frac{ie}{\hbar}\int_j^k\bm A\cdot d\bm\ell}
 \equiv e^{i\lambda_{kj}},
 \label{eq:sm-cot-transfer}
\end{equation}
where the full path from $j$ to $k$ runs through $M$.
Pair creation instead transfers two electrons from the Andreev dot to the
normal dots. Both tunneling amplitudes are outward, so both inward
Peierls factors are conjugated. Together with the phase of the
Andreev-dot pair-creation term in Eq.~\eqref{eq:sm-abs}, they give
\begin{equation}
 e^{i\varphi_M}
 e^{+\frac{ie}{\hbar}\int_j^M\bm A\cdot d\bm\ell}
 e^{+\frac{ie}{\hbar}\int_k^M\bm A\cdot d\bm\ell}
 \equiv e^{i\theta_{kj}}.
 \label{eq:sm-pair-transfer}
\end{equation}
These scalar factors multiply the cotunneling and crossed-Andreev
amplitudes derived in the main text. Reversing the second arm integral
in Eq.~\eqref{eq:sm-pair-transfer}, we obtain the general phase formulas
\begin{align}
 \lambda_{kj}&=-\frac{e}{\hbar}\int_j^k\bm A\cdot d\bm\ell,
 \label{eq:sm-lambda}\\
 \theta_{kj}&=\varphi_M+\frac{e}{\hbar}
 \left(\int_j^M\bm A\cdot d\bm\ell
             -\int_M^k\bm A\cdot d\bm\ell\right).
 \label{eq:sm-theta}
\end{align}
Thus the hopping phase involves the sum of the two forward arm integrals,
whereas the pairing phase involves their difference and the condensate
phase at the Andreev dot. No midpoint assumption is needed for these
relations; below we specialize to straight bonds with
$\bm r_M=(\bm r_j+\bm r_k)/2$.

In the notation of the main-text spinless model $H_{\mathrm{cp}}$, the
orbital field makes the bondwise substitutions
\begin{equation}
 t\longmapsto t e^{i\lambda_{kj}},\qquad
 \Delta_0\longmapsto\Delta_0 e^{i\theta_{kj}}.
 \label{eq:sm-substitutions}
\end{equation}
The first substitution applies to the coefficient of
$-c_{\bm k}^\dagger c_{\bm j}$, and the second to the coefficient of
$c_{\bm k}^\dagger c_{\bm j}^\dagger$, retaining the intrinsic factor
$1$ for a positive $x$ bond and $-i$ for a positive $y$ bond. Here and
below, we use the chirality $\eta=-1$ (InAs) of the main text; for
$\eta=+1$, replace $-i$ by $+i$. Explicitly,
\begin{align}
 H_{\mathrm{cp}}={}&-t\sum_{\langle\bm j,\bm k\rangle}
 \left(e^{i\lambda_{\bm k\bm j}}c_{\bm k}^\dagger c_{\bm j}
       +\mathrm{H.c.}\right)
 -\mu\sum_{\bm j}c_{\bm j}^\dagger c_{\bm j}\nonumber\\
 &+\Delta_0\sum_{\bm j}\left[
 \left(e^{i\theta_{\bm j+\hat{\bm x},\bm j}}
            c_{\bm j+\hat{\bm x}}^\dagger
       -i e^{i\theta_{\bm j+\hat{\bm y},\bm j}}
            c_{\bm j+\hat{\bm y}}^\dagger\right)c_{\bm j}^\dagger
       +\mathrm{H.c.}\right].
 \label{eq:sm-lattice}
\end{align}
Here each nearest-neighbor bond is counted once, oriented in the positive
$x$ or $y$ direction, and $c_{\bm j}\equiv c_{\bm j\uparrow}$ as in the
main text. We distinguish the effective pairing strength $\Delta_0>0$
from the Andreev-dot pairing $\Delta=|\Delta|e^{i\varphi_M}$; its phase
$\varphi_M$ is already included in $\theta_{kj}$.

As a check, the gauge transformation $\bm A\mapsto\bm A+\nabla\chi$,
$\varphi_M\mapsto\varphi_M-2e\chi_M/\hbar$ gives
\begin{equation}
 \lambda_{kj}\mapsto\lambda_{kj}-\frac{e}{\hbar}(\chi_k-\chi_j),
 \qquad
 \theta_{kj}\mapsto\theta_{kj}-\frac{e}{\hbar}(\chi_k+\chi_j).
 \label{eq:sm-gauge}
\end{equation}
The Andreev-dot gauge phase cancels from both expressions. The remaining
endpoint phases cancel those of the electron operators,
$c_{\bm j}\mapsto e^{-ie\chi_{\bm j}/\hbar}c_{\bm j}$, in
Eq.~\eqref{eq:sm-lattice}.

\section{Homogeneous background field in Landau gauge}
\label{sec:sm-uniform}

The normal dots occupy $\bm r_j=a(n_x,n_y)$ with integers $n_x,n_y$.
The superconducting wires lie on the lines $x=(n_x+\tfrac12)a$ and
$y=(n_y+\tfrac12)a$, and pass through the Andreev dots. We take a prescribed
uniform background field with
\begin{equation}
 B_{\mathrm{bg}}a^2=m\frac{h}{e}=2m\Phi_0,\qquad
 \Phi_0=\frac{h}{2e},\qquad m\in\mathbb Z,
 \quad
 \bm A_{\mathrm{bg}}=(0,B_{\mathrm{bg}}x).
 \label{eq:sm-background}
\end{equation}
The horizontal hopping phase is zero. For a vertical bond starting at
$x=n_xa$, Eq.~\eqref{eq:sm-lambda} gives
\begin{equation}
 \lambda_x^{\mathrm{bg}}=0,\qquad
 \lambda_y^{\mathrm{bg}}=-\frac{eB_{\mathrm{bg}}n_xa^2}{\hbar}
                       =-2\pi m n_x.
 \label{eq:sm-uniform-hopping}
\end{equation}
Thus every background hopping factor is unity. Along each straight bond,
the tangential vector potential is constant, so the two forward
half-bond integrals in Eq.~\eqref{eq:sm-theta} are equal. The explicit
orbital contribution to the pairing phase cancels, leaving
$\theta_{kj}^{\mathrm{bg}}=\varphi_M^{\mathrm{bg}}$.

To determine this remaining phase, we impose zero supercurrent along the
superconducting wires. The tangential gauge-invariant phase gradient then
vanishes:
\begin{equation}
 \partial_\parallel\varphi^{\mathrm{bg}}
       =-\frac{2e}{\hbar}A_{\mathrm{bg},\parallel}.
 \label{eq:sm-zero-current}
\end{equation}
This condition is compatible with a single-valued order parameter,
which requires an integer multiple of $\Phi_0$ through each
superconducting loop: a counterclockwise circuit of any square
accumulates
\begin{equation}
 \Delta_\square\varphi^{\mathrm{bg}}
 =-\frac{2e}{\hbar}B_{\mathrm{bg}}a^2=-4\pi m.
 \label{eq:sm-background-winding}
\end{equation}
In Landau gauge, $\partial_x\varphi^{\mathrm{bg}}=0$ on horizontal wires,
whereas $\partial_y\varphi^{\mathrm{bg}}=-2eB_{\mathrm{bg}}x/\hbar$ on
vertical wires. Starting at $(-a/2,-a/2)$ with phase zero and following
the central square counterclockwise, the bottom edge has phase zero,
the right edge lowers the phase by $2\pi m$, the top edge has phase
$-2\pi m$, and the left edge lowers it by another $2\pi m$.
Every vertex consequently has the same complex phase.

On the full network we may choose a local phase branch that is zero on
each horizontal wire. On a vertical segment at
$x_v=(n_x+\tfrac12)a$, with lower endpoint $y_-=(n_y-\tfrac12)a$, it is
\begin{equation}
 \varphi^{\mathrm{bg}}(x_v,y)
 =-\frac{2eB_{\mathrm{bg}}}{\hbar}x_v(y-y_-),
 \qquad y_-\le y\le y_-+a.
 \label{eq:sm-background-network}
\end{equation}
The upper endpoint differs from the lower one by
$-2\pi m(2n_x+1)$, a multiple of $2\pi$, so the complex order parameter
is continuous at every junction; the $2\pi$ jumps of the real phase
record the winding around the loops.

A horizontal normal-dot bond has its Andreev dot at
$((n_x+\tfrac12)a,n_y a)$, halfway along a vertical superconducting
segment. A vertical normal-dot bond has its Andreev dot on a horizontal
superconducting wire. Hence
\begin{equation}
 \varphi_{M_x}^{\mathrm{bg}}=-\pi m(2n_x+1),\qquad
 \varphi_{M_y}^{\mathrm{bg}}=0.
 \label{eq:sm-background-pairing}
\end{equation}
The background therefore leaves $t$ unchanged and gives the bondwise
substitution
\begin{equation}
 \Delta_0\longmapsto
 \begin{cases}
 (-1)^m\Delta_0,&\text{positive $x$ bond},\\
 \Delta_0,&\text{positive $y$ bond}.
 \end{cases}
 \label{eq:sm-background-substitution}
\end{equation}
The $y$-bond pair-creation coefficient in $H_{\mathrm{cp}}$ retains its
intrinsic factor $-i$.

\section{An additional localized flux in a symmetric gauge}
\label{sec:sm-vortex}

We now add flux $s\Phi_0$, with $s=\pm1$, spread uniformly over the
central superconducting square. Define
\begin{equation}
 g(u)=\operatorname{clip}(u,-a/2,a/2)=
 \begin{cases}
 -a/2,&u<-a/2,\\
 u,&-a/2\le u\le a/2,\\
 a/2,&u>a/2.
 \end{cases}
 \label{eq:sm-clip}
\end{equation}
Its derivative is one inside the central interval and zero outside.
A fourfold-symmetric vector potential for the extra field is
\begin{equation}
 \bm A_v=\frac{s\Phi_0}{2a^2}
 \begin{pmatrix}-g'(x)g(y)\\g(x)g'(y)\end{pmatrix},\qquad
 B_v=\partial_x A_{v,y}-\partial_y A_{v,x}
    =\frac{s\Phi_0}{a^2}g'(x)g'(y).
 \label{eq:sm-vortex-gauge}
\end{equation}
Inside the square this reduces to $\bm A_v=(B_v/2)(-y,x)$;
outside, its curl vanishes. We use the mixed gauge
$\bm A=\bm A_{\mathrm{bg}}+\bm A_v$, retaining Landau gauge for the
background. The zero-current condition is linear, so we also choose
$\varphi=\varphi^{\mathrm{bg}}+\varphi^v$. Each contribution separately
has integer superconducting flux through every network loop.

\subsection{Condensate phase on the superconducting network}

We fix the extra phase by $\varphi^v(a/2,0)=0$. Integrating
$d\varphi^v=-(2e/\hbar)\bm A_v\cdot d\bm\ell$ from this Andreev dot to
the upper-right corner gives
\begin{equation}
 \varphi^v(a/2,a/2)
 =-\frac{2e}{\hbar}\frac{s\Phi_0}{4a}\frac a2
 =-\frac{s\pi}{4}.
 \label{eq:sm-eighth-turn}
\end{equation}
Each successive half-edge contributes the same phase change. The four
central Andreev-dot phases, ordered counterclockwise from the positive
$x$ axis, are $0$, $-s\pi/2$, $-s\pi$, and $-3s\pi/2$, with total winding
$-2\pi s$. These signs follow from the electron charge $-e$ and the
pair-creation convention in Eq.~\eqref{eq:sm-lattice}.

The solution extends to every superconducting wire. On a horizontal
wire at $y_h=(n_y+\tfrac12)a$, and on a vertical wire at
$x_v=(n_x+\tfrac12)a$, convenient local branches are
\begin{align}
 \varphi_H^v(x,y_h)&=
 \begin{cases}
 \displaystyle \frac{s\pi}{2a}g(x)-\frac{s\pi}{2},&y_h>0,\\[3pt]
 \displaystyle-\frac{s\pi}{2a}g(x)-\frac{3s\pi}{2},&y_h<0,
 \end{cases}\label{eq:sm-horizontal-condensate}\\
 \varphi_V^v(x_v,y)&=
 \begin{cases}
 \displaystyle-\frac{s\pi}{2a}g(y),&x_v>0,\\[3pt]
 \displaystyle \frac{s\pi}{2a}g(y)-s\pi,&x_v<0.
 \end{cases}\label{eq:sm-vertical-condensate}
\end{align}
The horizontal and vertical expressions agree as complex phases at
their intersections. A lift that decreases by $2\pi s$ on a
counterclockwise circuit is obtained by subtracting $2\pi s$ from the
right-hand vertical branch below the $x$ axis. The positive $x$ axis
then forms the branch seam. In particular, the extra condensate phase
is zero at every Andreev dot on the positive $x$ axis.

Away from the central strips, the tangential vector potential vanishes
and the condensate phase stays constant along the wires in each
quadrant. At Andreev-dot positions the coordinates are integer or
half-integer multiples of $a$, so $g(u)=(a/2)\operatorname{sgn}(u)$
there, with $\operatorname{sgn}(0)=0$. Introducing the auxiliary angle
\begin{equation}
 \vartheta_r=\arg\bigl[\operatorname{sgn}(x_r)
                         +i\operatorname{sgn}(y_r)\bigr],
 \qquad \bm r_r\ne0,
 \label{eq:sm-sign-angle}
\end{equation}
Eqs.~\eqref{eq:sm-horizontal-condensate} and
\eqref{eq:sm-vertical-condensate} give
\begin{equation}
 e^{i\varphi_M^v}=e^{-is\vartheta_M}.
 \label{eq:sm-condensate-angle}
\end{equation}
The angle $\vartheta$ takes values in steps of $\pi/4$ and is not the
polar angle.

\subsection{Hopping phases and pairing-arm corrections}

Integrating Eq.~\eqref{eq:sm-vortex-gauge} along the normal-dot bonds
gives the extra hopping phases
\begin{align}
 \lambda_x^v(x,y)&=\frac{s\pi}{2a^2}g(y)
                         [g(x+a)-g(x)],\label{eq:sm-vortex-hopping-x}\\
 \lambda_y^v(x,y)&=-\frac{s\pi}{2a^2}g(x)
                         [g(y+a)-g(y)].\label{eq:sm-vortex-hopping-y}
\end{align}
For the pairing, the difference of the forward half-bond integrals
instead yields
\begin{align}
 \theta_x^v(x,y)&=\varphi^v(x+a/2,y)
 -\frac{s\pi}{2a^2}g(y)
 [2g(x+a/2)-g(x)-g(x+a)],\label{eq:sm-vortex-pairing-x}\\
 \theta_y^v(x,y)&=\varphi^v(x,y+a/2)
 +\frac{s\pi}{2a^2}g(x)
 [2g(y+a/2)-g(y)-g(y+a)].\label{eq:sm-vortex-pairing-y}
\end{align}
Unlike the uniform-field contribution, these arm corrections generally
survive because the two half-bonds need not sample the same tangential
vector potential. On the four bonds attached to the central normal
dot they vanish, as do the extra hopping phases. On the bond $(0,a)\to(a,a)$,
for example, the Andreev-dot phase is $-s\pi/4$ and the arm correction is
$-s\pi/8$, giving $\theta^v=-3s\pi/8$.

\subsection{Compact expression for the extra pairing phase}

The discrete geometry permits a formula involving only the endpoint
coordinates. Let $\bm u_r=(\operatorname{sgn}x_r,
\operatorname{sgn}y_r)$ and define
\begin{equation}
 C_{jk}=u_{M,x}(u_{j,y}+u_{k,y})
                -u_{M,y}(u_{j,x}+u_{k,x}).
 \label{eq:sm-cross-signs}
\end{equation}
Substituting $g=(a/2)\operatorname{sgn}$ into
Eqs.~\eqref{eq:sm-vortex-pairing-x} and
\eqref{eq:sm-vortex-pairing-y} combines both orientations into
\begin{equation}
 \theta_{kj}^v=\varphi_M^v-\frac{s\pi}{8}C_{jk}.
 \label{eq:sm-discrete-arm}
\end{equation}
For noncentral endpoints, the sign-coordinate angles can be lifted
locally, on the same branch as $\vartheta_M$, so that
$|\vartheta_j-\vartheta_k|\le\pi/4$. A nearest-neighbor
step either leaves their sign directions unchanged or connects an axis
to an adjacent quadrant. The corresponding angular relation is
\begin{equation}
 \vartheta_j+\vartheta_k-2\vartheta_M
        =\frac{\pi}{4}C_{jk}.
 \label{eq:sm-angle-relation}
\end{equation}
The arm correction therefore shifts the Andreev-dot phase to the average
endpoint angle. Choosing the compatible representative
$\varphi_M^v=-s\vartheta_M$, we obtain
\begin{equation}
 \theta_{kj}^v=-s\vartheta_M
 -\frac{s}{2}(\vartheta_j+\vartheta_k-2\vartheta_M)
 =-\frac{s}{2}(\vartheta_j+\vartheta_k).
 \label{eq:sm-half-angle}
\end{equation}

Define unit complex numbers at the noncentral normal dots by
\begin{equation}
 z_r^{(s)}=e^{-is\vartheta_r}
 =\frac{\operatorname{sgn}(x_r)-is\operatorname{sgn}(y_r)}
 {\sqrt{\operatorname{sgn}^2(x_r)+\operatorname{sgn}^2(y_r)}}.
 \label{eq:sm-z}
\end{equation}
The half-angle identity gives
\begin{equation}
 z_j^{(s)}+z_k^{(s)}
 =2\cos\!\left(\frac{\vartheta_j-\vartheta_k}{2}\right)
       e^{-is(\vartheta_j+\vartheta_k)/2}.
 \label{eq:sm-phasor-sum}
\end{equation}
The cosine is positive on the local branches just specified.
Normalizing the sum consequently removes its magnitude and retains
the pairing phase. The result extends to the central bonds by setting
$z_0^{(s)}=0$: their arm correction vanishes, and the Andreev dot has the
same sign direction as the noncentral endpoint. Thus, on every
nearest-neighbor bond,
\begin{equation}
 F_{kj}^{(s)}\equiv e^{i\theta_{kj}^v}
 =\frac{z_j^{(s)}+z_k^{(s)}}{|z_j^{(s)}+z_k^{(s)}|},
 \qquad z_0^{(s)}=0.
 \label{eq:sm-compact-vortex}
\end{equation}
The normalization fixes the magnitude to $\Delta_0$ and avoids the
branch ambiguity of averaging two principal angles.

Combining the extra flux with the Landau-gauge background gives the
substitutions in the effective model:
\begin{equation}
 t\longmapsto t e^{i\lambda_{kj}^v},\qquad
 \Delta_0\longmapsto\Delta_0
 \begin{cases}
 (-1)^m F_{kj}^{(s)},&k=j+\hat{\bm x},\\
 F_{kj}^{(s)},&k=j+\hat{\bm y}.
 \end{cases}
 \label{eq:sm-final-coefficients}
\end{equation}
The minus sign of the hopping term and the intrinsic pairing factors
$1$ and $-i$ remain as in Eq.~\eqref{eq:sm-lattice}.
Reversing the extra flux conjugates its phase factors:
$\lambda_{kj}^{v,(-)}=-\lambda_{kj}^{v,(+)}$ and
$F_{kj}^{(-)}=(F_{kj}^{(+)})^*$, while the background and intrinsic
chirality factors remain fixed. These expressions apply at any lattice
size for the specified square flux profile and midpoint Andreev dots.
\section{Numerical details}
\label{sec:sm-numerics}

This section complements the description of the numerical simulations in the End Matter of the main text, with $\tau_{x,y,z}$ the Pauli matrices in the local Nambu basis $(c_{\bm j},c_{\bm j}^\dag)$.

\subsection{Majorana wave functions}
To separate vortex core and edge Majorana wave functions, we work within the lowest particle-hole symmetric pair at energies $\pm E_0$ with wave functions $\psi_+$ and $\psi_-$.
We choose their phases such that $\psi_-=\mathcal C\psi_+$, where $\mathcal C=\tau_xK$ is the particle-hole operator.
Then, the combinations
\begin{equation}\label{eq:numerical-majorana-basis}
 \chi_1=\frac{\psi_++\psi_-}{\sqrt2},\qquad
 \chi_2=\frac{\psi_+-\psi_-}{i\sqrt2}
\end{equation}
are orthonormal and Majorana modes, $\mathcal C\chi_a=\chi_a$.
Rotations within this basis preserve the Majorana nature and allow us to optimize the spatial separation.
We project the perimeter weight onto this subspace,
\begin{equation}\label{eq:numerical-perimeter}
 W_{ab}=\sum_{\bm j\in\partial\Omega}\chi_{a,\bm j}^\dag\chi_{b,\bm j},
 \qquad a,b\in\{1,2\},
\end{equation}
where $\partial\Omega$ contains the outermost dots of the $L\times L$ lattice.
For a normalized real combination $\chi=q_1\chi_1+q_2\chi_2$, the perimeter weight is $q^TWq$.
In the eigenbasis of the real symmetric matrix $W$, this is a weighted average of its eigenvalues.
Its smallest and largest eigenvalues therefore give the minimum and maximum perimeter weights, and the corresponding eigenvectors specify the two respective Majorana wave functions.
We use these combinations to identify the core and edge modes, respectively.

\subsection{Conductances}
From the scattering matrix $S(E)$ computed with Kwant~\cite{groth2014}, we obtain the scattering probabilities $T_{\ell\ell'}^{\alpha\beta}(E)$ from lead $\ell'$ to lead $\ell$, with $\ell,\ell'\in\{L,T,R\}$ and outgoing and incoming quasiparticle types $\alpha,\beta\in\{e,h\}$, respectively.
We obtain the charge and thermal conductances from these probabilities using the scattering formalism reviewed in Ref.~\cite{waintal2024}.
We define the differential conductance as $G_{\ell\ell'}=\partial I_\ell/\partial V_{\ell'}$, with the charge current $I_\ell$ counted positive when flowing from lead $\ell$ into the island and lead electrochemical potentials $\mu_\ell-\mu_S=-eV_\ell$.
A finite bias $V_{\ell'}=V$ is applied to lead $\ell'$, while all other leads and the superconducting condensate remain grounded.
At zero temperature this reduces to
\begin{equation}\label{eq:numerical-charge}
 G_{\ell\ell'}(V)=\frac{e^2}{h}
 \bigl[\delta_{\ell\ell'}-T_{\ell\ell'}^{ee}(-eV)+T_{\ell\ell'}^{he}(-eV)\bigr].
\end{equation}
We also consider the thermal conductance at temperature $T$ in the absence of a voltage bias.
In linear response, it is given by
\begin{equation}\label{eq:numerical-heat}
 G_{\ell\ell'}^{\mathrm{th}}=
 \frac{1}{hT}\int_0^{2 t_\ell} dE\,E^2
 \biggl(-\frac{\partial f_T}{\partial E}\biggr)
 \biggl[2\delta_{\ell\ell'}-\sum_{\alpha,\beta}T_{\ell\ell'}^{\alpha\beta}(E)\biggr],
\end{equation}
where $f_T(E)=[e^{E/(k_BT)}+1]^{-1}$.
The integral is evaluated adaptively using SciPy~\cite{virtanen2020}.
Thermal currents are also defined positive when flowing from the lead into the system, hence $-G_{\ell\ell'}^{\mathrm{th}}$ measures heat transmitted from $\ell'$ to $\ell$ (for $\ell\ne\ell'$).